\def\fouru{4U1915-05}
\def\ergs{ergs s$^{-1}$}
\def\ctsspcu{Counts s$^{-1}$ PCU$^{-1}$}

\def   \ni {\noindent}

\def   \ssk {\vskip  5truept}

\def   \bsk {\vskip 15truept}
 
\def   \newpage {\vfill\eject}
\def   \newline {\hfil\break}

\documentstyle[epsfig]{article}
\begin{document}

\hsize 5truein
\vsize 8truein
\font\abstract=cmr8
\font\keywords=cmr8
\font\caption=cmr8
\font\references=cmr8
\font\text=cmr10
\font\affiliation=cmssi10
\font\author=cmss10
\font\mc=cmss8
\font\title=cmssbx10 scaled\magstep2
\font\alcit=cmti7 scaled\magstephalf
\font\alcin=cmr6 
\font\ita=cmti8
\font\mma=cmr8
\def\ref{\par\noindent\hangindent 15pt}
\null


\title{\ni DISCOVERY OF HIGH FREQUENCY QUASI-PERIODIC OSCILLATIONS IN 4U1915-05}

\bsk \bsk
\author{\ni D. Barret$^{1}$, L. Boirin$^{1}$, J. F. Olive$^{1}$, 
J.E. Grindlay$^{2}$, P.F. Bloser$^{2}$, J.H. Swank$^{3}$ \&
A.P. Smale$^{3}$}
\bsk
\affiliation{1) Centre d'Etude Spatiale des Rayonnements, CNRS-UPS, 9 Avenue du Colonel Roche, 31028 Toulouse Cedex 04, France}

\affiliation{2) Harvard Smithsonian Center for Astrophysics, 60 Garden Street, Cambridge, MA 02138, USA}

\affiliation{3) Laboratory for High Energy Astrophysics, NASA Goddard Space Flight Center, Greenbelt, MD 20771, USA}
\bsk
\baselineskip = 12pt

\abstract{ABSTRACT \ni
The type I X-ray burster and dipper 4U1915-05 (also known as
XB1916-053) was monitored by the {\it Rossi X-ray Timing Explorer}
between February and October, 1996.  The source was observed in
various spectral states; the highest luminosity state (L$_{\rm
X}\sim1.5\times10^{37}$\ergs, 10 kpc, 1-20 keV) is associated with a
soft spectrum, whereas for the lower luminosity state (down to $\sim
5\times10^{36}$\ergs) the spectrum is significantly harder. Using the
high time resolution data provided by the Proportional Counter Array
(PCA), we have discovered High-Frequency Quasi-Periodic Oscillations
(HFQPOs) in the persistent X-ray emission of \fouru~while its
luminosity was $\sim 8\times 10^{36}$\ergs. The QPO frequency ranges
from 600 Hz up to $\sim 1000$ Hz, with typical Full Width at Half
Maximum (FWHM) $\sim 50-100$ Hz and Root Mean Squared (RMS) values
$\sim 15$\%. In addition, by using the ``shift and add'' technique, we
have detected a twin HFQPO ($5.5 \sigma$ level) separated from the
main peak by $\sim 355$ Hz. \fouru~is the eighth Atoll source
displaying simultaneous twin HFQPOs. Based on current knowledge of
HFQPO sources, our observations suggest that \fouru~might contain a
2.8 (or 5.6) millisecond rotating neutron star.}
\bsk
\baselineskip = 12pt

\bsk
\baselineskip = 12pt


\text{\ni 1. INTRODUCTION
\ssk
\ni     
HFQPOs ranging from $\sim 300$ to $\sim 1200$ Hz have now been
detected from over 15 neutron star Low Mass X-ray Binaries (LMXBs)
(Van der Klis 1998). The origin of these HFQPOs remains unclear, but
they are most likely to be produced close to the neutron star. The
frequency of the QPOs is generally positively correlated with the
inferred mass accretion rate. They are seen over a limited range of
luminosity; for instance in Atoll sources, HFQPOs are generally
detected in the so-called ``Island'' state, and on the lower branch of
the ``Banana'' state.  In most sources, twin HFQPOs are detected with
a frequency separation in the range 250-350 Hz (Van der Klis 1998). In
four cases, coherent oscillations during X-ray bursts have been also
observed at a frequency which is equal or half the value of the
separation between the twin peaks.

\newpage
\begin{center}
\begin{table}[t]
\begin{center}
\begin{tabular}{|llccllc|}
\hline
Date & PCUs & T (s) & R & $\nu$ (Hz)  & FWHM & RMS (\%)   \\
\hline
\hline
1996:05:16 & 3 E& 7600 & 37.89 & 1006.4$^{+16.2}_{-16.6}$ & 79.8$^{+37.2}_{-29.9}$ & $16.4\pm1.7$ \\
\hline
1996:05:23 &4-5 E& 7400 & 33.22 & 835.1$^{+14.7}_{-14.2}$ & 93.4$^{+35.8}_{-28.1}$ & $16.8\pm1.3$ \\
segment      & & 3200 & 32.58 & 818.9$^{+11.4}_{-10.9}$& 51.2$^{+23.1}_{-18.4}$ & $13.9\pm1.5$ \\
segment   & & 3200 & 32.96 & 845.5$^{+15.9}_{-15.9}$& 78.4$^{+36.2}_{-25.5}$ & $15.9\pm1.6$ \\ 
\hline
1996:06:01 &5 E& 5600 & 35.51 & 932.9$^{+14.9}_{-14.9}$ & 111.2$^{+34.7}_{-29.2}$ & $16.8\pm1.1$  \\
segment        &            & 3000 & 35.20 & 911.3$^{+16.8}_{-16.4}$ & 103.2$^{+43.8}_{-28.2}$ & $18.3\pm1.5$  \\
segment       &            & 2600 & 35.88 & 953.1$^{+13.8}_{-13.8}$ & 71.7$^{+24.7}_{-20.4}$ & $14.5\pm1.4$  \\
\hline
1996:06:01 &5 G& 2800 & 36.86 & 935.2$^{+13.3}_{-13.2}$ & 89.0$^{+30.1}_{-25.2}$ & $16.6\pm1.2$  \\
\hline
1996:09:06 &5 E& 8800 & 33.63 &  869.4$^{+16.2}_{-15.7}$ &147.5$^{+34.2}_{-28.8}$ & $19.7\pm1.1$\\
segment                & &2400 & 34.05 &  844.1$^{+11.5}_{-10.8}$ & 73.0$^{+22.4}_{-19.5}$ & $20.2\pm1.5$ \\
segment                    & &2200 & 34.56 &  888.6$^{+32.4}_{-30.2}$ &172.8$^{+68.4}_{-66.6}$ & $22.5\pm2.0$  \\
segment       & &2000 & 34.22 &  923.8$^{+23.2}_{-22.7}$ &125.4$^{+56.6}_{-45.0}$ & $19.5\pm1.7$  \\
\hline
1996:10:29 &5 E& 9000 & 35.11 & 604.4$^{+8.5}_{-8.5}$ &  45.3$^{+17.3}_{-15.9}$ & $11.7\pm1.1$  \\ 
segment         &              & 2800 & 37.40 & 610.2$^{+4.1}_{-4.1}$ &  20.7$^{+8.5}_{-6.4}$ & $11.9\pm1.2$ \\
\hline
\end{tabular}
\end{center}
\caption{TABLE 1. The properties of the HFQPOs detected from \fouru. 
The date of the observation, the number of PCA units (PCUs) operating
during the observation, the data used (E or G), the exposure time (T)
in seconds, the count rate (R) in units of \ctsspcu, the centroid
frequency of the QPO (fitted by a gaussian), its FWHM in Hz, and its
RMS in \% in the 5-30 keV range are listed. E stands for Event mode
data (122 $\mu s$), while G means {\it GoodXenon data} ($0.95\mu s$).}
\end{table} 
\end{center}

In the simplest interpretation, the burst oscillation reveals the spin
frequency of the neutron star, the higher of the two HFQPOs is
associated with a disk keplerian frequency, whereas a beat frequency
mechanism accounts for the lower of the two QPO peaks. However, this
interpretation has recently been called into question by new data,
which showed that the frequency separation of the twin peaks was not
constant (e.g. 4U1608-522, Mendez et al. 1998a), or even that the
burst frequency was not equal to the frequency difference (4U1636-53,
Mendez et al. 1998b).

In this paper, we report on the discovery of HFQPOs from the LMXB
\fouru; a type I X-ray burster and dipper with a 50 minute
orbital period, and a probable Atoll source (Yoshida 1992). A more
detailed paper, describing both the correlated timing and spectral
behavior, as well as the noise properties of the source (which
confirmed its Atoll nature) will appear elsewhere (Boirin et
al. 1998).
\bsk
\ni 2. DISCOVERY OF HFQPOs FROM 4U1915-05
\ssk
\ni 
\fouru~was observed by RXTE for about 140 ksec in 19 snapshot 
observations covering from February to October 1996. The main data
sets were separated in time by roughly one month. The net source count
rate varied from $\sim 15$ to $\sim 70$ \ctsspcu~(5-30 keV). The
source was thus observed in different spectral/intensity states as
inferred from the hardness intensity diagram shown in Fig. 1. The
global trend is that in the highest count rate regime, the spectrum is
soft and cuts off around 10 keV, whereas for the lowest count rates,
the spectrum is much harder and takes approximately a power law shape
in the PCA energy range with no observable cutoffs below $\sim 20$
keV. A paper devoted to the spectral analysis of the persistent
emission of the source will be published elsewhere (Bloser et
al. 1998).

\begin{figure}[t]
\vspace*{-1.5cm}
\centerline{\psfig{figure=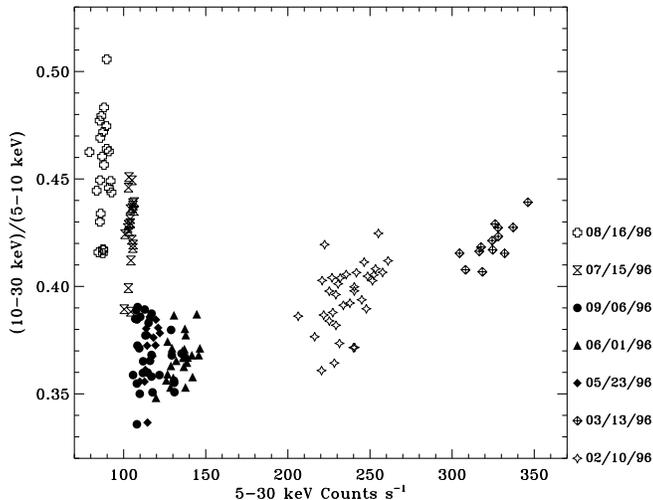,height=10cm}}
\vspace*{-1.cm}
\caption{FIGURE 1. Hardness intensity diagram of \fouru. The 
X-axis represents the background subtracted count rates in the 5-30
keV band, while the Y-axis is the ratio of the counts in the 10-30 keV
band to the counts in the 5-10 keV band. Only data recorded with the 5
PCU units are shown. Filled symbols are for observations in which
HFQPOs were detected (see below). The dates of the observation are
listed on the right handside of the plot.}
\end{figure}

To study the $\sim$ 100-1200 Hz variability of the X-ray emission, we
used both the 122 $\mu$s Event mode and the so-called {\it GoodXenon}
(0.95$\mu$s) high time resolution PCA data. Each continuous set
(bursts filtered out) was divided into segments of 4096 bins lasting
250 $\mu$s. Fast Fourier Transform (FFT) were computed on each
segment. 200 of these FFT were then averaged to obtain a final Power
Density Spectrum (PDS). The analysis has been carried out in the 5-30
keV band, where we have found that the signal to noise ratio of the
HFQPOs was maximum.

We have detected, above the $5\sigma$ significance level, HFQPOs from
\fouru~in 5 observations (see Table 1). The statistical significance of
the signals ranges from 5 to $\sim 10\sigma$. In some observations,
(e.g. September 6th, 1996), the HFQPOs could also be detected in short
($\sim 2000-3000$ seconds) segments of the observation. The three
strongest HFQPO signals are shown in Fig. 2. Unfortunately, the
weakness of the signals does not allow to study the RMS of the HFQPOs
as a function of energy. Nevertheless in the May 23rd observation,
there may be a positive correlation between the HFQPO RMS and energy
(see Boirin et al. 1998 for more details). Similar correlations have
been observed in other sources (Van der Klis 1998).

\begin{center}
\begin{figure}[!h]
\vspace*{-0.5cm}
\centerline{\psfig{figure=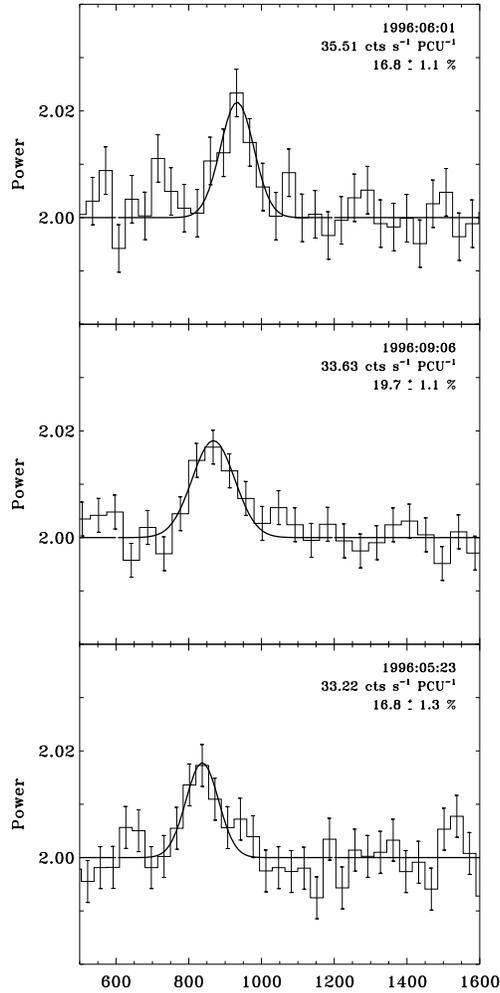,height=14cm}}
\caption{FIGURE 2. The three strongest HFQPOs detected. The date 
of the observation, the 5-30 keV count rate (per PCU and before
background subtraction), and the RMS of the signal are shown at the
top right of the windows. Note the relative weakness of the signals
compared to other sources. Note also the positive correlation between
the QPO frequency and the count rate.}
\end{figure}
\end{center}

Looking at Fig. 1, one can see that HFQPOs were not detected in
observations with the largest count rates. We have derived a $3\sigma$
upper limit of $\sim 8$\% on the RMS of a signal above 200 Hz of
FWHM=100 Hz (5-30 keV range). HFQPOs were not detected either in the
lowest count rate regime. Unfortunately our upper limit of $\sim 13$\%
on the RMS is not really stringent. A correlation between the QPO
frequencies and the source count rate which is common to most sources
is also seen from \fouru~as shown in Fig. 3.

\begin{center}
\begin{figure}[!h]
\vspace*{-1.5cm}
\centerline{\psfig{figure=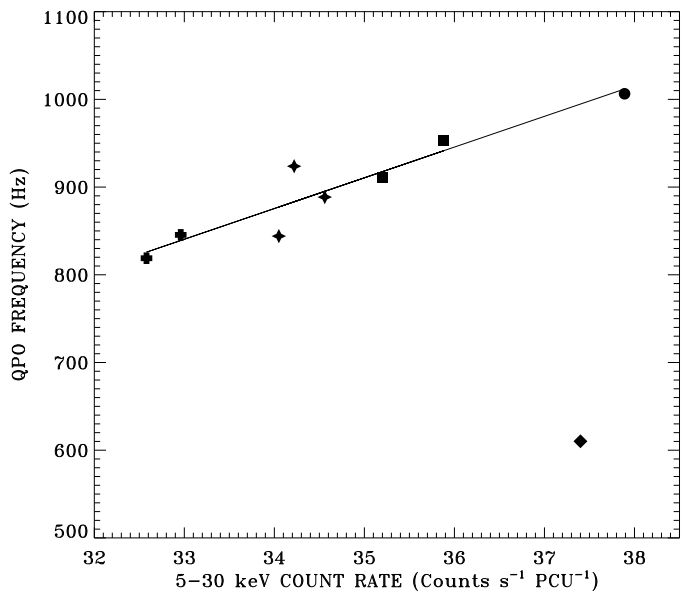,height=10cm}}
\vspace*{-0.5cm}
\caption{FIGURE 3. Correlation between the QPO frequency and the 
count rate listed in Table 1.  Note that for the upper data points,
the count rate varies by at most 20\% whereas the QPO frequency varies
from 800 to 1000 Hz. Note also that for the same count rates, the
HFQPO can either be found around 600 Hz or 1000 Hz.}
\end{figure}
\end{center}

On May 16th and June 1st, a signal (below our confidence threshold)
has been detected at 655 Hz (3.0$\sigma$) and 556 Hz ($4.6\sigma$)
respectively, implying a frequency separation of $\sim 350$ Hz. We
have thus applied the ``shift and add'' technique (Mendez et
al. 1998c) to the May 16th and June 1st observations. As expected, a
significant twin HFQPO ($5.5 \sigma$) shows up at 650 $\pm 5$ Hz
(FWHM=$25\pm10$ Hz, RMS$\sim 10.0$\%), whereas the main peak is
located at 1005 Hz. This yields a frequency separation of $\sim 355$
Hz between the two peaks (Fig. 4).

\begin{center}
\begin{figure}[!t]
\vspace*{-0.5cm}
\centerline{\psfig{figure=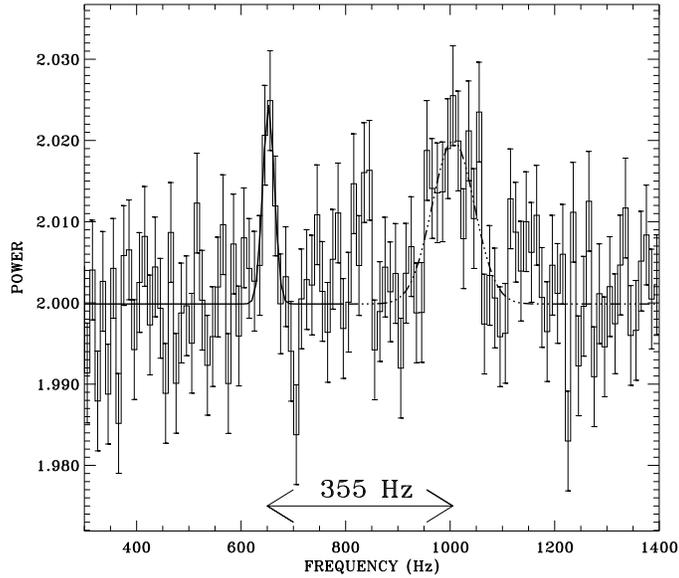,height=10cm}}
\vspace*{-0.5cm}
\caption{FIGURE 4. The twin HFQPOs in \fouru. The separation between 
the two peaks is 355 Hz in the range of frequency separation observed
in similar systems. The significance of the two peaks above the mean
Poisson level is larger than $5 \sigma$.}
\end{figure}
\end{center}

\vspace*{-1cm}
\bsk
\ni 3. CONCLUSIONS 
\ssk
\ni 
4U 1915-05 thus joins the class of Atoll sources displaying HFQPOs. In
\fouru, the HFQPOs are detected over a limited range of count rates,
when the source X-ray luminosity is intermediate in our data set (i.e.
$\sim 8\times10^{36}$\ergs), most likely on the so-called lower branch
of the ``Banana'' state. HFQPOs were not detected in the higher
luminosity states. HFQPOs are not detected either at the lowest count
rates, but our upper limits are not very stringent. We have also found
evidence for a twin HFQPO, separated from the main peak by about 355
Hz. Thus \fouru~becomes the eighth Atoll source to display
simultaneous twin HFQPOs. Further observations of the source are
needed to better determine the properties of its HFQPOs; in particular
a more complete sampling of its luminosity/spectral states is
required. These observations are also needed to determine what is the
actual spin frequency of the neutron star. In the framework of the
simple interpretation discussed above, and existing observations of
similar systems, our data suggest that it could be $\sim 355$ Hz, or
half that value. If this is confirmed, this would mean that
\fouru~contains a 2.8 (or 5.6) millisecond rotating neutron star.
\bsk
\baselineskip = 12pt
{\abstract \ni ACKNOWLEDGMENTS

We thank E. Ford, M. Van der Klis, W. Kluzniak for useful discussions
during the meeting.
 }

\bsk
\baselineskip = 12pt


{\references \ni REFERENCES
\ssk
\ref Bloser, P. et al., 1998, ApJ, in preparation
\ref Boirin, L. et al., 1998, A\&A, in preparation
\ref Grindlay J.E. et al., 1988, ApJ, 334, L25 
\ref Mendez M. et al., 1998a, ApJL, 505, L23
\ref Mendez M. et al., 1998b, ApJ, 506, L117
\ref Mendez M. et al., 1998c, ApJ, 494, L65
\ref Van der Klis, M., 1998, in AIP Conf. Proc 431, 361
\ref Yoshida K., 1992, PhD Thesis, Tokyo University
}                      

\end{document}